\documentclass[reprint, superscriptaddress, aps, showpacs, floatfix, prl]{revtex4-1}

\pdfoutput=1

\usepackage{graphicx}	

\usepackage{dcolumn}

\usepackage{bm}					
\usepackage{amsmath,amsbsy,amsfonts,revsymb}

\def\wn/{\,cm$^{-1}$}
\def\area/{\,cm$^{-2}$}
\def\BFO/{BiFeO${_3}$}
\def\cubic/{$_\mathrm{c}$}
\def\DM/{Dzyaloshinskii-Moriya}
\def\TN/{T_{\rm N}}
\def\Tc/{T_{\rm c}}
\def \vS {{\bf S}}
\def \vR {{\bf R}}

\def \vm {{\bf m}}

\def \mb {\mu_{\rm B}}

\def \xp {{\bf x}^{\prime }}
\def \yp {{\bf y}^{\prime }}
\def \zp {{\bf z}^{\prime }}
\def \xx {{\bf x}}
\def \yy {{\bf y}}
\def \zz {{\bf z}}

\def \zq {z^{\prime}}

\begin{document}

\title{Terahertz spectroscopy of spin waves in multiferroic BiFeO$_3$ in high magnetic fields}

\author{U. Nagel}
\email{urmas.nagel@kbfi.ee}
\affiliation{National Institute of Chemical Physics and Biophysics, Akadeemia tee 23, 12618 Tallinn, Estonia} 

\author{Randy S. Fishman}
\affiliation{Materials Science and Technology Division, Oak Ridge National Laboratory, Oak Ridge, Tennessee 37831, USA}

\author{T. Katuwal}
\altaffiliation{On leave from 
Department of Physics, 
Trichandra College, Tribhuvan University, Kathmandu, Nepal}
\affiliation{National Institute of Chemical Physics and Biophysics, Akadeemia tee 23, 12618 Tallinn, Estonia} 

\author{H. Engelkamp}
\affiliation{
Radboud University Nijmegen, Institute for Molecules and Materials, High Field Magnet Laboratory, Toernooiveld 7, 6525 ED  Nijmegen, the Netherlands}

\author{D. Talbayev}
\affiliation{Department of Physics, Tulane University, 5032 Percival Stern Hall, New Orleans, LA 70118, USA}

\author{Hee Taek Yi}
\affiliation{Rutgers Center for Emergent Materials and Department of Physics and Astronomy, Rutgers University, 136 Frelinghuysen Rd., Piscataway, New Jersey 08854, USA}

\author{S.-W. Cheong}
\affiliation{Rutgers Center for Emergent Materials and Department of Physics and Astronomy, Rutgers University, 136 Frelinghuysen Rd., Piscataway, New Jersey 08854, USA}

\author{T. R{\~o}{\~o}m}
\affiliation{National Institute of Chemical Physics and Biophysics, Akadeemia tee 23, 12618 Tallinn, Estonia} 

\date{\today}

\begin{abstract}
We have studied the magnetic field dependence of far-infrared active magnetic modes in a single ferroelectric domain \BFO/ crystal at low temperature.
The modes soften close to the critical field of 18.8\,T along the [001] (pseudocubic) axis, where the cycloidal structure changes to the homogeneous canted antiferromagnetic state and a new strong mode with linear field dependence appears that persists at least up to 31\,T. 
A microscopic model that includes two \DM/ interactions and easy-axis anisotropy describes closely both the zero-field spectroscopic modes as well as their splitting and evolution in a magnetic field.
The good agreement of theory with experiment suggests that the proposed model provides the foundation for future technological applications of this multiferroic material.
\end{abstract}


\maketitle

\begin{figure}
%
\includegraphics[width=0.3\textwidth]{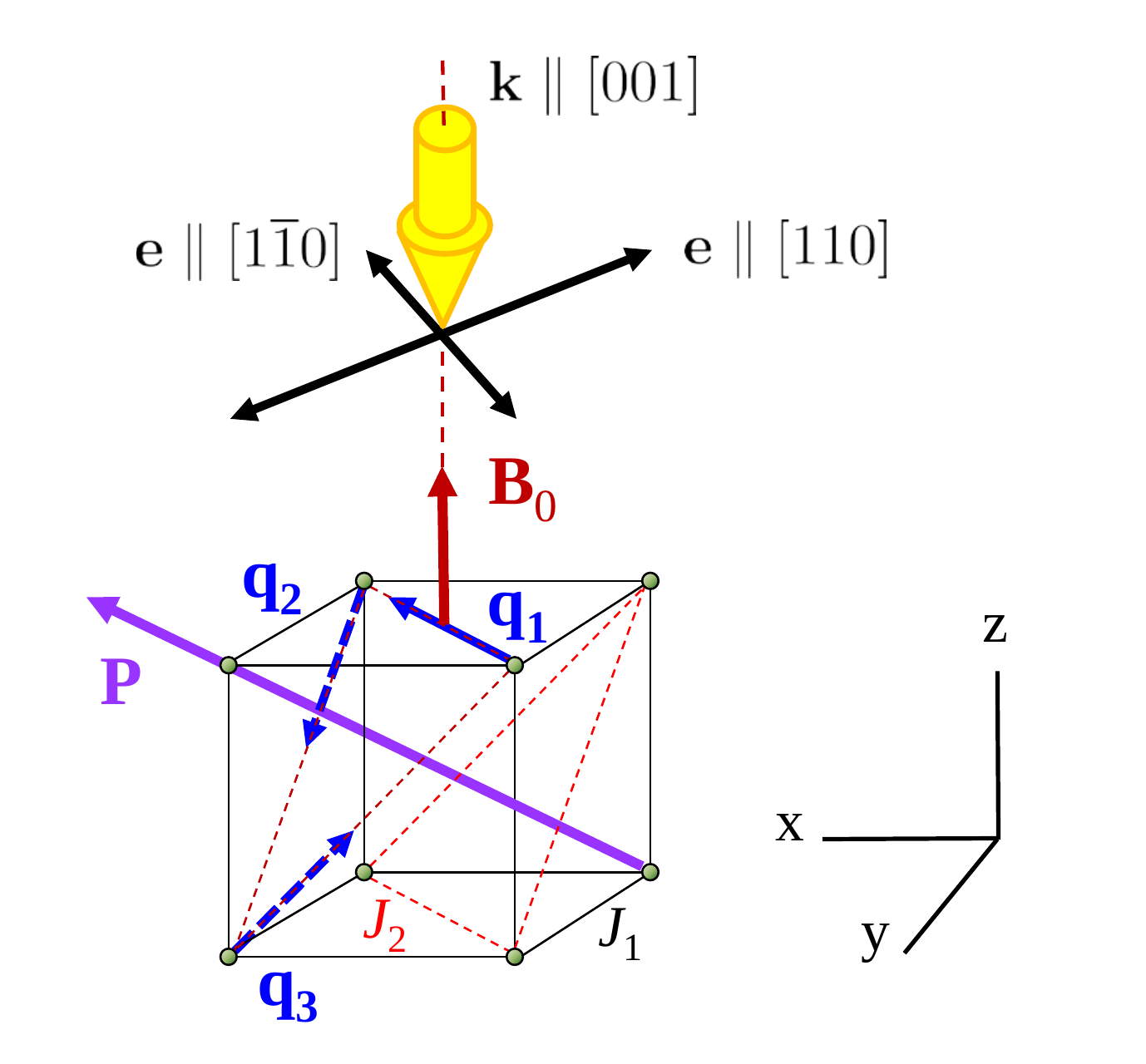}
\caption{\label{fig:model} 
(color online)
Pseudocubic unit cell of \BFO/ showing the positions of  Fe ions, the ferroelectric polarization $\mathbf{P}$, three equivalent directions of the cycloidal ordering vector $\mathbf{q}_k$, the applied static magnetic field $\mathbf{B}_0 \parallel [001]$, and the wave vector of incident light $\mathbf{k}$ together with the electric field ($\mathbf{e}$) component of light in two orthogonal polarizations that were used in the experiment.
$J_1$ and $J_2$ are the nearest- and next-nearest neighbor exchange interactions.
} 
\end{figure}

\begin{figure}
%
\includegraphics[width=0.49\textwidth]{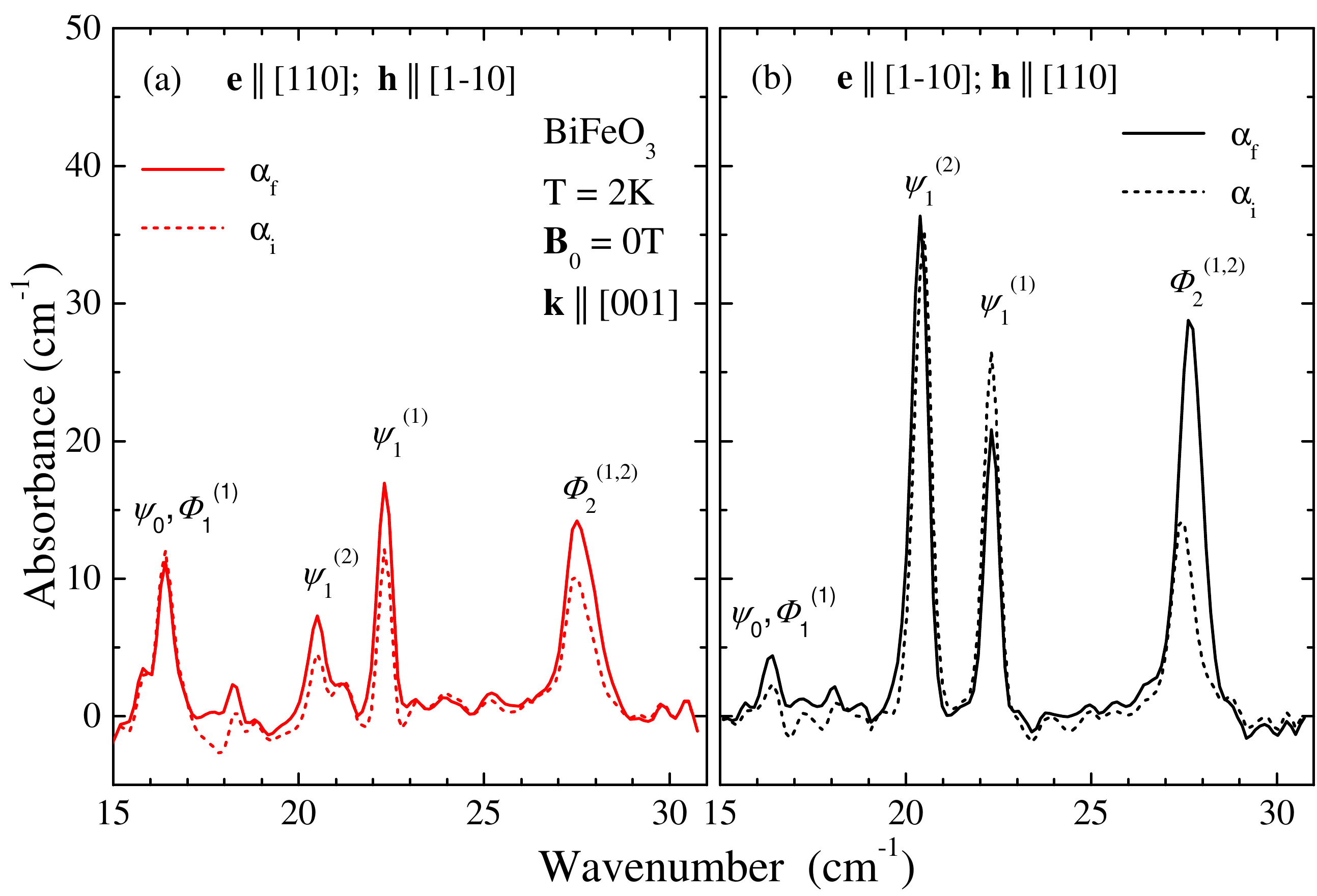}
\caption{\label{fig:HistZFieldLinesPolDep} 
(color online)
Absorbance spectra of spin wave modes in zero field in $\mathbf{e}\parallel [110]$ polarization (a) and in $\mathbf{e}\parallel [1\bar{1}0]$ polarization (b). Solid curves, $\alpha_{F}$, were measured after applying the high field $\mathbf{B}_0 \geq 21$\,T.
Dotted curves are initial absorbance spectra $\alpha_{I}$ of the zero field cooled sample.
}
\end{figure}


Due to the coupling between electric and magnetic properties, multiferroic materials are among the most important yet discovered.  
With a multiferroic material used as a storage medium, information can be written electrically and then read magnetically without 
Joule heating\cite{Eerenstein2006}.  Hence, applications of a room-temperature multiferroic would radically transform the magnetic storage industry.
Because it is the only known room-temperature multiferroic, \BFO/ continues to attract intense interest.

Although its ferroelectric transition temperature\cite{Teague1970} $\Tc/ \approx 1100$\,K is much higher than its N\'eel transition temperature\cite{Sosnowska1982, Lebeugle2008, Lee2008} $\TN/ \approx 640$\,K, the appearance of a long-wavelength cycloid\cite{Sosnowska1982, Ramazanoglu2011, Herrero-Albillos2010, Sosnowska2011} with a period of 62 nm enhances the ferroelectric polarization below $\TN/ $.  
The induced polarization has been used to switch between magnetic domains with an applied electric field\cite{Lebeugle2008,Lee2008,Lee2008prb}.  

Progress in understanding the microscopic interactions in \BFO/ has been greatly accelerated by the recent availability of single crystals for both elastic and inelastic neutron-scattering measurements.  
By fitting the spin wave frequencies above a few meV, recent measurements\cite{Jeong2012, Matsuda2012} have determined the antiferromagnetic (AFM) nearest-neighbor and 
next-nearest neighbor exchanges $J_1 \approx -4.5$ meV and $J_2=-0.2$ meV.
In the presence of strain\cite{Bai2005}, non-magnetic impurities\cite{Chen2012}, or a magnetic field\cite{Tokunaga2010JPSJ, Park2011} above $B_c \approx 19$ T, those exchange interactions produce a G-type antiferromagnet with ferromagnetic alignment of the $S=5/2$ Fe$^{3+}$ spins within each hexagonal plane, (111) in cubic notation.

Below $B_c$, the magnetic order in \BFO/ is created by the much smaller anisotropy and Dzyaloshinskii-Moriya interactions.
Neutron scattering is typically used to determine the weak interactions that produce a complex spin state.
Because the wavelength $a/\sqrt{2}\delta $ of \BFO/ is so large, however, inelastic neutron-scattering measurements cannot resolve the cycloid satellite peaks at ${\bf q}=(2\pi /a)(0.5\pm \delta , 0.5, 0.5 \mp \delta )$, on either side of the AFM wavevector ${\bf Q}_0 = (\pi /a)(1,1,1)$, where $a \approx 3.96$\,\AA\   is the pseudocubic lattice constant.
Below 5 meV, inelastic measurements at ${\bf q}_0$ reveal four broad peaks,  each of which can be roughly assigned to one or more of the spin wave branches averaged over the first Brillouin zone\cite{Matsuda2012, Fishman2012}.
By contrast, THz spectroscopy\cite{Talbayev2011, Huvonen2009} provides very precise measurements for the optically-active spin wave frequencies at the cycloid wavevector ${\bf q}$.

Symmetry allows three possible directions of the cycloidal ordering vector $\mathbf{q}_k\,||\,\left\lbrace[1,\bar{1},0],[0,1,\bar{1}],[\bar{1},0,1] \right\rbrace \perp \mathbf{P}$, see Fig.\,\ref{fig:model}. 
The spins of a cycloid $\mathbf{q}_k$ are in the plane determined by $\mathbf{P}$ and $\mathbf{q}_k$.
The cycloidal order in \BFO/ is induced by a weak \DM/ interaction that couples spins along $\mathbf{q}_k$ with coupling $\mathbf{D} \perp \mathbf{P}$ and $\mathbf{D} \perp \mathbf{q}_k$\cite{Kadomtseva2004}.
Another \DM/ like  interaction $\mathbf{D}'\parallel\mathbf{P}\parallel [111]$ couples spins on $[111]$ direction.
It is induced by magnetoelectric coupling and cants spins out of the cycloid plane\cite{Kadomtseva2004,Ederer2005,Albrecht2010}.
The ferromagnetic ordering of this canted moment has been verified by a neutron scattering experiment\cite{Ramazanoglu2011prl}.
High resolution neutron scattering shows that the magnetic ground state ordering in \BFO/ does not change in zero field on cooling from 300\,K to 4\,K\cite{Przenioslo2006,Albillos2010,Ramazanoglu2011}.

Single ion anisotropy $K$ along the easy axis [111] introduces anharmonicity\cite{Kadomtseva2004,Kadomtseva2006}, but in zero magnetic field the cycloid is only slightly anharmonic\cite{Ramazanoglu2011,Sosnowska2011}.
An external magnetic field contributes to the effective single ion anisotropy\cite{Tehranchi1997} and induces a metamagnetic transition\cite{Kadomtseva2004} at the critical field $B_c \approx 19$\,T, where the cycloidal order changes to a collinear AFM spin order\cite{Ohoyama2011}.
The unwinding of the cycloid reduces the electric polarization\cite{Park2011} and creates a small macroscopic spontaneous magnetization induced by $\mathbf{D}'\parallel [111]$\cite{Kadomtseva2004,Park2011}.

The frequencies of magnetic modes are sensitive to anisotropic magnetic interactions and these interactions are important to understand the microscopic models behind the magnetoelectric coupling in multiferroics.
The eigenspectrum of \BFO/ cycloids was calculated by de Sousa and Moore\cite{DeSousa2008}, with the addition of single-ion easy-axis anisotropy by Fishman et al.\cite{Fishman2012, Fishman2013PRB} in 0\,T and in applied electric field by Rovillain et al.\cite{Rovillain2010}.
Spectroscopic techniques that measure the eigenspectrum of magnetic modes are valuable tools, especially if they can be combined with external fields that compete with internal fields. 
The Raman work demonstrated that the Raman-active spin wave frequencies depend strongly on applied electric field\cite{Rovillain2010}.
Most of the INS\cite{Jeong2012,Matsuda2012}, Raman\cite{Cazayous2008,Rovillain2009} and THz\cite{Komandin2010,Talbayev2011} spectroscopy studies on \BFO/ were in zero applied field.
The high field ESR was done in magnetic fields up to 25\,T, but was limited to frequencies lower than the main cycloid modes and one of the AFM modes. 

In this Letter, we present THz absorption spectra of a \BFO/ single crystal at low temperature and follow the magnetic field dependence of cycloid excitations until the cycloidal order is destroyed in high magnetic field and replaced by a canted AFM order.
We show that the proposed microscopic model in addition to describing the frequencies of the cycloid in zero field, also predicts the splitting and evolution of the spin wave modes with magnetic field\cite{Fishman2012, Fishman2013PRB}.
Due to mode mixing, all of the spin wave mode become optically active in magnetic field.
The close agreement between predictions and measurements suggests that the proposed model can provide the foundation for future work on \BFO/. 


In a magnetic field ${\bf H}=H\vm $ along $\vm $, the spin state and spin wave excitations of \BFO/ are evaluated from the Hamiltonian
\begin{eqnarray}
&&{\cal H} = -J_1\sum_{\langle i,j\rangle }\vS_i\cdot \vS_j -J_2\sum_{\langle i,j \rangle'} \vS_i\cdot \vS_j
-K\sum_i (\vS_i \cdot \zp )^2
\nonumber \\
&&-{D\, \sum}_{\vR_j=\vR_i + a(\xx -\zz )} \,\yp \cdot (\vS_i\times\vS_j) \nonumber \\
&& - {D'\, \sum}_{\vR_j=\vR_i + a\xx, a\yy, a\zz } \, (-1)^{R_{i\zq } /c} \,\zp \cdot  (\vS_i\times\vS_j)\nonumber \\
&& - 2\mb H \sum_i \vS_i \cdot \vm .
\label{Ham}
\end{eqnarray}
Here $\zp \,||\, [1,1,1]$, $\xp \,||\, \mathbf{q}_k$, and $\yp = \zp \times \xp $, where $k=1,2,3$ are the indexes of the three cycloids that are symmetry-equivalent in zero field.
While the nearest- and next-nearest neighbor exchange interactions $J_1=-4.5$ meV and $J_2=-0.2$ meV can be obtained from the spin wave dispersion between 5.5\,meV and 72\,meV\cite{Jeong2012, Matsuda2012} 
the small interactions $D=0.107$\,meV, $D'=0.054$\,meV, and $K = 0.0035$\,meV that control the cycloid\cite{Fishman2013PRB} are obtained from the THz spectra below 5.5\,meV (44.3\wn/), measured in zero magnetic field.
For a given set of interaction parameters and magnetic field, the spin state of \BFO/ is obtained by minimizing the energy $E=\langle {\cal H}\rangle $.


The (001) face single crystal \BFO/ sample was grown using a Bi$_2$O$_3$ flux\cite{Choi2009}.
It has a thickness of 0.37\,mm and it contains a single ferroelectric domain,  $\mathbf{P}\parallel[111]$ axis in Fig.\,\ref{fig:model}, checked by an optical rectification experiment\cite{Talbayev2008}.

The sample was zero field cooled and spectra were measured in Faraday configuration with the magnetic field along the [001] axis. 
Up to 12\,T spectra were measured at 4\,K in Tallinn with a SPS-200 Martin-Puplett spectrometer from Sciencetech Inc. and a 0.3\,K bolometer\cite{room2004NaVa} using a spectral resolution of 0.2\wn/.
Spectra from 12\,T up to 31\,T were measured in Nijmegen High Field Magnet Laboratory at 2\,K using a Bruker IFS 113v spectrometer and a 1.6\,K silicon bolometer and spectral resolution of 0.43\wn/; the spectra were averaged for 15 minutes at each field.
There was a linear polarizer in front of the sample to control the polarization of light.

\begin{figure}
%
\includegraphics[width=0.49\textwidth]{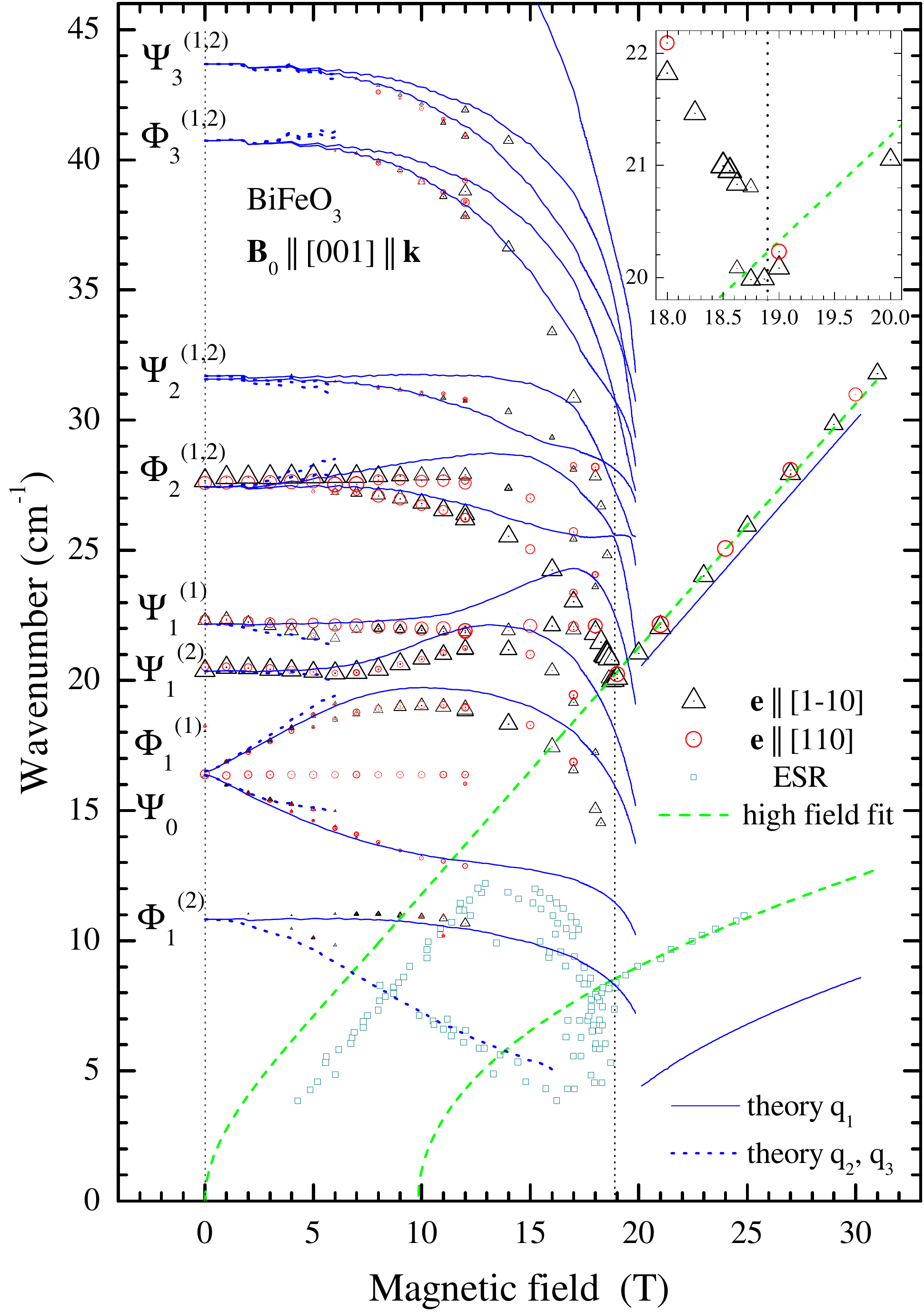}
\caption{\label{fig:FarBFOlinepos} 
(color online)
Magnetic field dependence of spin wave modes in the THz absorption spectrum of \BFO/ at low temperature.
The areas of triangles and circles are proportional to the absorption line areas.
Vertical dashed line at $B_c = 18.8$\,T marks the metamagnetic transition.
Green dashed lines are the fit of our data and ESR data\cite{Ruette2004} (squares) above 19\,T to a model from Ref.\,\onlinecite{Ruette2004}.
Blue solid lines are calculated modes of cycloid $q_1$.
Dotted blue lines are calculated modes of cycloids $q_2$ and $q_3$, shown only below 6\,T where corresponding excitations are observed in measured spectra; the lowest energy mode is shown also for higher fields since there is a matching excitation in ESR data.
The theoretically predicted metamagnetic transition is at 20\,T for cycloid $\mathbf{q}_1$ and at 16.5\,T for cycloids $\mathbf{q}_2$ and $\mathbf{q}_3$.
} 
\end{figure}

We measured absorbance spectra in magnetic field with the reference spectrum in zero field.
This method gave excellent spectra of magnetic field dependent lines.
From the differential absorbance spectra in fields above 21\,T (after applying 30\,T) we extracted the zero field absorption lines, solid curves in Fig.\,\ref{fig:HistZFieldLinesPolDep} and fitted them.
The fit results were added to the measured differential spectra in magnetic fields.
The result, absolute absorbance spectra in fields, is shown in\cite{PRL2013suppl}  and fitted line positions and areas are shown in Fig.\,\ref{fig:FarBFOlinepos}.

A change in the zero field spectra was observed after applying high field at low temperature, Fig.\,\ref{fig:HistZFieldLinesPolDep}.
The zero field spectrum stayed the same after applying high field again or in the opposite direction.
The initial zero field line intensities $\alpha_\mathrm{I}$, measured on the zero field cooled sample, were recovered after warming the sample to 300\,K.
This is evidence that different magnetic domains exist.
We found that the change in the zero field spectra, as measured after applying magnetic fields at low temperature, occurs already by 12\,T and higher fields that destroy the cycloid do not change the zero field lines any more\cite{PRL2013suppl}.
The calculation shows that for $\mathbf{B}_0 \parallel$\,[001] the $\mathbf{q}_1$ cycloid where the field is mostly perpendicular to the cycloid plane has a lower energy than cycloids $\mathbf{q}_2$ and $\mathbf{q}_3$\cite{PRL2013suppl}.
However, some fraction of $\mathbf{q}_2$ and $\mathbf{q}_3$ domains follow the magnetic field without hysteresis below 5\,T as discussed below.
This suggests that crystal imperfections acts as a barrier to maintain the dominant  $\mathbf{q}_1$ domain after the field has been removed.
Clearly there is a thermally activated hysteresis, but in this study we concentrate on the low temperature spectra, that are measured after the sample had been in high field $\geq 12$\,T.

Detailed field dependence of mode frequencies and areas is presented in Fig.\,\ref{fig:FarBFOlinepos}.
The three main modes, $\Psi_{1}^{(2)}$, $\Psi_{1}^{(1)}$ and $\Phi_{2}^{(1,2)}$ change only slightly with increasing magnetic field until about 5\,T is reached, where a discontinuity of several mode frequencies and a smooth change in the slope of the $\Psi_{1}^{(1)}$ mode is observed.
These changes are associated with the change in the magnetic domain structure, where modes of the $q_2$ and $q_3$ domains (blue dotted lines) are depopulated and only the modes of the $q_1$ domain (blue solid lines) remain observable in higher fields.
To reflect this behavior we have cut off the predicted mode frequencies of domains $q_2$ and $q_3$ above 6\,T.

Modes soften before reaching the metamagnetic transition at $B_c$, except $\Psi_{1}^{(1)}$ which seems to merge with the softening $\Phi_{2}^{(1,2)}$ at about 18\,T.
There is an intriguing possibility that close to the transition between 18.6 and 18.8\,T, see inset to Fig.\ref{fig:FarBFOlinepos}, one of the cycloid resonances, $\Psi_{1}^{(1)}$ or $\Phi_{2}^{(1,2)}$ as labeled in zero field, coexists with the AFM resonance.
This means that in a narrow field interval the spin structure supports both, cycloidal and AFM modes.
The coexistence of two phases is ruled out since the metamagnetic transition in \BFO/ is neither the first order phase transition nor similar to a spin flop transition in ordinary antiferromagnets\cite{Tehranchi1997} and also we did not observe any hysteresis effects between 18 and 19\,T as reported earlier\cite{Ruette2004}.
In THz spectra there is only one resonance line above 18.8\,T and we assign this value to the critical field $B_c$ of the metamagnetic transition in \BFO/ at 2\,K and $\mathbf{B}_0\parallel [001]$.
However, the THz spectra do not show any anomalies at 10\,T seen by optical measurements\cite{Xu2009bfo}.

The AFM resonances in \BFO/ can also be described by a phenomenological theory\cite{Ruette2004} and we use it to fit the ESR\cite{Ruette2004} and THz data in the homogeneous canted antiferromagnetic state, see dashed lines in Fig.\,\ref{fig:FarBFOlinepos}.
The fit gives the following parameters: gyromagnetic ratio $\gamma = (1.72 \pm 0.01) \times 10^{7}\,\mathrm{rad}(\mathrm{s}\,\mathrm{Oe})^{-1}$, $K_u/\chi_{\perp} = (1.06 \pm 0.02) \times 10^{10}\, \mathrm{erg\,cm^{-3}}$, $ H_{DM}=(105 \pm 2)\mathrm{kOe}$.
$K_u$ is the energy density of the uniaxial magnetic anisotropy and $\chi_{\perp}$ is the susceptibility perpendicular to the AFM vector, the difference of magnetizations of two sublattices of a G-type antiferromagnet.
$H_{DM}$ is the \DM/ field associated with $\mathbf{D}'$.

We get $K_u = 6.2 \times 10^{5}$\,erg\,cm$^{-3}$, using $\chi_{\perp}=5.8\times 10^{-5}$ from the high field magnetization measurement\cite{Tokunaga2010JPSJ}.
The value we get for the same quantity from our microscopic theory is $\langle KS_z^2 \rangle = 5.6 \times 10^{5}$\,erg\,cm$^{-3}$, where $K=0.0035$\,meV and $S_z=5/2$.
The canted moment as estimated from the AFM resonance spectra, $\chi_\perp H_{DM}=0.041\mu_B/\mathrm{Fe}$, should be compared to $ 0.03\mu_B$ derived from the extrapolation of the high field  magnetization to the zero field\cite{Tokunaga2010JPSJ}.
Thus, the parameters derived from the phenomenological model are very close to the values used in the microscopic theory.

Considering the microscopic theory, the agreement between the measured and predicted mode frequencies in Fig.\ref{fig:FarBFOlinepos} is remarkable.
In agreement with predictions, $\Psi_0$ and $\Phi_1^{(1)}$ are slightly lower in domain $q_1$ than in domains $q_2$ and $q_3$.  
The predicted splitting of $\Phi_2^{(1,2)}$ is clearly seen in Fig.\ref{fig:FarBFOlinepos}.
Also in agreement with predictions, $\Psi_1^{(1)}$ is slightly lower in domains $q_2$ and $q_3$ than in domain $q_1$.  
The only feature that remains unexplained by our model is the field-independent mode observed at about 16.5 cm$^{-1}$, midway between $\Phi_1^{(1)}$ and $\Psi_0$ which becomes too weak to be detected in the Nijmegen setup and thus cannot be followed until it disappears at $\mathbf{B}_c=18.8$\,T.
Notice that several modes in Fig.\ref{fig:FarBFOlinepos} only become optically active in magnetic field.
Recall that in our microscopic model we use the same interaction parameters that were previously obtained for zero field\cite{Fishman2013PRB}.
Therefore, it is not surprising that the quantitative agreement with measurements, although quite good, is not perfect.
In particular, the lower frequency AFM mode in the canted phase is predicted about 4 cm$^{-1}$ lower than measured by ESR.


To conclude, the close agreement between the predicted and observed spin wave frequencies in magnetic field leaves no doubt about the origin of those modes.  
This agreement suggests that the present model, with \DM/ interactions along $\yp $ and $\zp$ and easy-axis anisotropy along $\zp $, can provide the foundation for future studies on \BFO/ and may lay the groundwork for its eventual technological applications.
Our work demonstrates that in addition to electric field\cite{Lee2008} the control of magnetic domains with magnetic field is possible.

We acknowledges conversations with Nobuo Furukawa, Masaaki Matsuda, Shin Miyahara, Satoshi Okamoto, and Rogerio de Sousa.
We acknowledge support by the Estonian Ministry of Education and Research grant No. SF0690029s09, Estonian Science Foundation grant Nos. ETF8170, ETF8703 and ERMOS67, and by EuroMAgNET under the EU Contract No. 228043.
Work at Rutgers was supported by NSF-DMR-1104484.
RSF acknowledges support by the U.S. Department of Energy, Office of Basic Energy Sciences, Materials Sciences and Engineering Division.


\begin{thebibliography}{39}%
\makeatletter
\providecommand \@ifxundefined [1]{%
 \@ifx{#1\undefined}
}%
\providecommand \@ifnum [1]{%
 \ifnum #1\expandafter \@firstoftwo
 \else \expandafter \@secondoftwo
 \fi
}%
\providecommand \@ifx [1]{%
 \ifx #1\expandafter \@firstoftwo
 \else \expandafter \@secondoftwo
 \fi
}%
\providecommand \natexlab [1]{#1}%
\providecommand \enquote  [1]{``#1''}%
\providecommand \bibnamefont  [1]{#1}%
\providecommand \bibfnamefont [1]{#1}%
\providecommand \citenamefont [1]{#1}%
\providecommand \href@noop [0]{\@secondoftwo}%
\providecommand \href [0]{\begingroup \@sanitize@url \@href}%
\providecommand \@href[1]{\@@startlink{#1}\@@href}%
\providecommand \@@href[1]{\endgroup#1\@@endlink}%
\providecommand \@sanitize@url [0]{\catcode `\\12\catcode `\$12\catcode
  `\&12\catcode `\#12\catcode `\^12\catcode `\_12\catcode `\%12\relax}%
\providecommand \@@startlink[1]{}%
\providecommand \@@endlink[0]{}%
\providecommand \url  [0]{\begingroup\@sanitize@url \@url }%
\providecommand \@url [1]{\endgroup\@href {#1}{\urlprefix }}%
\providecommand \urlprefix  [0]{URL }%
\providecommand \Eprint [0]{\href }%
\providecommand \doibase [0]{http://dx.doi.org/}%
\providecommand \selectlanguage [0]{\@gobble}%
\providecommand \bibinfo  [0]{\@secondoftwo}%
\providecommand \bibfield  [0]{\@secondoftwo}%
\providecommand \translation [1]{[#1]}%
\providecommand \BibitemOpen [0]{}%
\providecommand \bibitemStop [0]{}%
\providecommand \bibitemNoStop [0]{.\EOS\space}%
\providecommand \EOS [0]{\spacefactor3000\relax}%
\providecommand \BibitemShut  [1]{\csname bibitem#1\endcsname}%
\let\auto@bib@innerbib\@empty
\bibitem [{\citenamefont {Eerenstein}\ \emph {et~al.}(2006)\citenamefont
  {Eerenstein}, \citenamefont {Mathur},\ and\ \citenamefont
  {Scott}}]{Eerenstein2006}%
  \BibitemOpen
  \bibfield  {author} {\bibinfo {author} {\bibfnamefont {W.}~\bibnamefont
  {Eerenstein}}, \bibinfo {author} {\bibfnamefont {N.~D.}\ \bibnamefont
  {Mathur}}, \ and\ \bibinfo {author} {\bibfnamefont {J.~F.}\ \bibnamefont
  {Scott}},\ }\href@noop {} {\bibfield  {journal} {\bibinfo  {journal}
  {Nature}\ }\textbf {\bibinfo {volume} {442}},\ \bibinfo {pages} {759}
  (\bibinfo {year} {2006})}\BibitemShut {NoStop}%
\bibitem [{\citenamefont {Teague}\ \emph {et~al.}(1970)\citenamefont {Teague},
  \citenamefont {Gerson},\ and\ \citenamefont {James}}]{Teague1970}%
  \BibitemOpen
  \bibfield  {author} {\bibinfo {author} {\bibfnamefont {J.~R.}\ \bibnamefont
  {Teague}}, \bibinfo {author} {\bibfnamefont {R.}~\bibnamefont {Gerson}}, \
  and\ \bibinfo {author} {\bibfnamefont {W.~J.}\ \bibnamefont {James}},\
  }\href@noop {} {\bibfield  {journal} {\bibinfo  {journal} {Solid State
  Commun.}\ }\textbf {\bibinfo {volume} {8}},\ \bibinfo {pages} {1073}
  (\bibinfo {year} {1970})}\BibitemShut {NoStop}%
\bibitem [{\citenamefont {Sosnowska}\ \emph {et~al.}(1982)\citenamefont
  {Sosnowska}, \citenamefont {Peterlin-Neumaier},\ and\ \citenamefont
  {Steichele}}]{Sosnowska1982}%
  \BibitemOpen
  \bibfield  {author} {\bibinfo {author} {\bibfnamefont {I.}~\bibnamefont
  {Sosnowska}}, \bibinfo {author} {\bibfnamefont {T.}~\bibnamefont
  {Peterlin-Neumaier}}, \ and\ \bibinfo {author} {\bibfnamefont
  {E.}~\bibnamefont {Steichele}},\ }\href@noop {} {\bibfield  {journal}
  {\bibinfo  {journal} {J. Phys. C: Solid State Phys.}\ }\textbf {\bibinfo
  {volume} {15}},\ \bibinfo {pages} {4835} (\bibinfo {year}
  {1982})}\BibitemShut {NoStop}%
\bibitem [{\citenamefont {Lebeugle}\ \emph {et~al.}(2008)\citenamefont
  {Lebeugle}, \citenamefont {Colson}, \citenamefont {Forget}, \citenamefont
  {Viret}, \citenamefont {Bataille},\ and\ \citenamefont
  {Goukasov}}]{Lebeugle2008}%
  \BibitemOpen
  \bibfield  {author} {\bibinfo {author} {\bibfnamefont {D.}~\bibnamefont
  {Lebeugle}}, \bibinfo {author} {\bibfnamefont {D.}~\bibnamefont {Colson}},
  \bibinfo {author} {\bibfnamefont {A.}~\bibnamefont {Forget}}, \bibinfo
  {author} {\bibfnamefont {M.}~\bibnamefont {Viret}}, \bibinfo {author}
  {\bibfnamefont {A.~M.}\ \bibnamefont {Bataille}}, \ and\ \bibinfo {author}
  {\bibfnamefont {A.}~\bibnamefont {Goukasov}},\ }\href {\doibase
  10.1103/PhysRevLett.100.227602} {\bibfield  {journal} {\bibinfo  {journal}
  {Phys. Rev. Lett.}\ }\textbf {\bibinfo {volume} {100}},\ \bibinfo {pages}
  {227602} (\bibinfo {year} {2008})}\BibitemShut {NoStop}%
\bibitem [{\citenamefont {Lee}\ \emph {et~al.}(2008{\natexlab{a}})\citenamefont
  {Lee}, \citenamefont {Ratcliff}, \citenamefont {Cheong},\ and\ \citenamefont
  {Kiryukhin}}]{Lee2008}%
  \BibitemOpen
  \bibfield  {author} {\bibinfo {author} {\bibfnamefont {S.}~\bibnamefont
  {Lee}}, \bibinfo {author} {\bibfnamefont {W.}~\bibnamefont {Ratcliff}},
  \bibinfo {author} {\bibfnamefont {S.-W.}\ \bibnamefont {Cheong}}, \ and\
  \bibinfo {author} {\bibfnamefont {V.}~\bibnamefont {Kiryukhin}},\ }\href@noop
  {} {\bibfield  {journal} {\bibinfo  {journal} {Appl. Phys. Lett.}\ }\textbf
  {\bibinfo {volume} {92}},\ \bibinfo {pages} {192906} (\bibinfo {year}
  {2008}{\natexlab{a}})}\BibitemShut {NoStop}%
\bibitem [{\citenamefont {Ramazanoglu}\ \emph
  {et~al.}(2011{\natexlab{a}})\citenamefont {Ramazanoglu}, \citenamefont
  {Ratcliff}, \citenamefont {Choi}, \citenamefont {Lee}, \citenamefont
  {Cheong},\ and\ \citenamefont {Kiryukhin}}]{Ramazanoglu2011}%
  \BibitemOpen
  \bibfield  {author} {\bibinfo {author} {\bibfnamefont {M.}~\bibnamefont
  {Ramazanoglu}}, \bibinfo {author} {\bibfnamefont {W.}~\bibnamefont
  {Ratcliff}}, \bibinfo {author} {\bibfnamefont {Y.~J.}\ \bibnamefont {Choi}},
  \bibinfo {author} {\bibfnamefont {S.}~\bibnamefont {Lee}}, \bibinfo {author}
  {\bibfnamefont {S.-W.}\ \bibnamefont {Cheong}}, \ and\ \bibinfo {author}
  {\bibfnamefont {V.}~\bibnamefont {Kiryukhin}},\ }\href {\doibase
  10.1103/PhysRevB.83.174434} {\bibfield  {journal} {\bibinfo  {journal} {Phys.
  Rev. B}\ }\textbf {\bibinfo {volume} {83}},\ \bibinfo {pages} {174434}
  (\bibinfo {year} {2011}{\natexlab{a}})}\BibitemShut {NoStop}%
\bibitem [{\citenamefont {Herrero-Albillos}\ \emph
  {et~al.}(2010{\natexlab{a}})\citenamefont {Herrero-Albillos}, \citenamefont
  {Catalan}, \citenamefont {Rodriguez-Velamazan}, \citenamefont {Viret},
  \citenamefont {Colson},\ and\ \citenamefont {Scott}}]{Herrero-Albillos2010}%
  \BibitemOpen
  \bibfield  {author} {\bibinfo {author} {\bibfnamefont {J.}~\bibnamefont
  {Herrero-Albillos}}, \bibinfo {author} {\bibfnamefont {G.}~\bibnamefont
  {Catalan}}, \bibinfo {author} {\bibfnamefont {J.~A.}\ \bibnamefont
  {Rodriguez-Velamazan}}, \bibinfo {author} {\bibfnamefont {M.}~\bibnamefont
  {Viret}}, \bibinfo {author} {\bibfnamefont {D.}~\bibnamefont {Colson}}, \
  and\ \bibinfo {author} {\bibfnamefont {J.~F.}\ \bibnamefont {Scott}},\
  }\href@noop {} {\bibfield  {journal} {\bibinfo  {journal} {J. Phys.: Condens.
  Matter}\ }\textbf {\bibinfo {volume} {22}},\ \bibinfo {pages} {256001}
  (\bibinfo {year} {2010}{\natexlab{a}})}\BibitemShut {NoStop}%
\bibitem [{\citenamefont {Sosnowska}\ and\ \citenamefont
  {Przenios{\l}o}(2011)}]{Sosnowska2011}%
  \BibitemOpen
  \bibfield  {author} {\bibinfo {author} {\bibfnamefont {I.}~\bibnamefont
  {Sosnowska}}\ and\ \bibinfo {author} {\bibfnamefont {R.}~\bibnamefont
  {Przenios{\l}o}},\ }\href {\doibase 10.1103/PhysRevB.84.144404} {\bibfield
  {journal} {\bibinfo  {journal} {Phys. Rev. B}\ }\textbf {\bibinfo {volume}
  {84}},\ \bibinfo {pages} {144404} (\bibinfo {year} {2011})}\BibitemShut
  {NoStop}%
\bibitem [{\citenamefont {Lee}\ \emph {et~al.}(2008{\natexlab{b}})\citenamefont
  {Lee}, \citenamefont {Choi}, \citenamefont {Ratcliff}, \citenamefont {Erwin},
  \citenamefont {Cheong},\ and\ \citenamefont {Kiryukhin}}]{Lee2008prb}%
  \BibitemOpen
  \bibfield  {author} {\bibinfo {author} {\bibfnamefont {S.}~\bibnamefont
  {Lee}}, \bibinfo {author} {\bibfnamefont {T.}~\bibnamefont {Choi}}, \bibinfo
  {author} {\bibfnamefont {W.}~\bibnamefont {Ratcliff}}, \bibinfo {author}
  {\bibfnamefont {R.}~\bibnamefont {Erwin}}, \bibinfo {author} {\bibfnamefont
  {S.-W.}\ \bibnamefont {Cheong}}, \ and\ \bibinfo {author} {\bibfnamefont
  {V.}~\bibnamefont {Kiryukhin}},\ }\href {\doibase 10.1103/PhysRevB.78.100101}
  {\bibfield  {journal} {\bibinfo  {journal} {Phys. Rev. B}\ }\textbf {\bibinfo
  {volume} {78}},\ \bibinfo {pages} {100101} (\bibinfo {year}
  {2008}{\natexlab{b}})}\BibitemShut {NoStop}%
\bibitem [{\citenamefont {Jeong}\ \emph {et~al.}(2012)\citenamefont {Jeong},
  \citenamefont {Goremychkin}, \citenamefont {Guidi}, \citenamefont {Nakajima},
  \citenamefont {Jeon}, \citenamefont {Kim}, \citenamefont {Furukawa},
  \citenamefont {Kim}, \citenamefont {Lee}, \citenamefont {Kiryukhin},
  \citenamefont {Cheong},\ and\ \citenamefont {Park}}]{Jeong2012}%
  \BibitemOpen
  \bibfield  {author} {\bibinfo {author} {\bibfnamefont {J.}~\bibnamefont
  {Jeong}}, \bibinfo {author} {\bibfnamefont {E.~A.}\ \bibnamefont
  {Goremychkin}}, \bibinfo {author} {\bibfnamefont {T.}~\bibnamefont {Guidi}},
  \bibinfo {author} {\bibfnamefont {K.}~\bibnamefont {Nakajima}}, \bibinfo
  {author} {\bibfnamefont {G.}~\bibnamefont {Jeon}}, \bibinfo {author}
  {\bibfnamefont {S.}~\bibnamefont {Kim}}, \bibinfo {author} {\bibfnamefont
  {S.}~\bibnamefont {Furukawa}}, \bibinfo {author} {\bibfnamefont
  {Y.}~\bibnamefont {Kim}}, \bibinfo {author} {\bibfnamefont {S.}~\bibnamefont
  {Lee}}, \bibinfo {author} {\bibfnamefont {V.}~\bibnamefont {Kiryukhin}},
  \bibinfo {author} {\bibfnamefont {S.-W.}\ \bibnamefont {Cheong}}, \ and\
  \bibinfo {author} {\bibfnamefont {J.}~\bibnamefont {Park}},\ }\href@noop {}
  {\bibfield  {journal} {\bibinfo  {journal} {Phys. Rev. Lett.}\ }\textbf
  {\bibinfo {volume} {108}} (\bibinfo {year} {2012})}\BibitemShut {NoStop}%
\bibitem [{\citenamefont {Matsuda}\ \emph {et~al.}(2012)\citenamefont
  {Matsuda}, \citenamefont {Fishman}, \citenamefont {Hong}, \citenamefont
  {Lee}, \citenamefont {Ushiyama}, \citenamefont {Yanagisawa}, \citenamefont
  {Tomioka},\ and\ \citenamefont {Ito}}]{Matsuda2012}%
  \BibitemOpen
  \bibfield  {author} {\bibinfo {author} {\bibfnamefont {M.}~\bibnamefont
  {Matsuda}}, \bibinfo {author} {\bibfnamefont {R.~S.}\ \bibnamefont
  {Fishman}}, \bibinfo {author} {\bibfnamefont {T.}~\bibnamefont {Hong}},
  \bibinfo {author} {\bibfnamefont {C.~H.}\ \bibnamefont {Lee}}, \bibinfo
  {author} {\bibfnamefont {T.}~\bibnamefont {Ushiyama}}, \bibinfo {author}
  {\bibfnamefont {Y.}~\bibnamefont {Yanagisawa}}, \bibinfo {author}
  {\bibfnamefont {Y.}~\bibnamefont {Tomioka}}, \ and\ \bibinfo {author}
  {\bibfnamefont {T.}~\bibnamefont {Ito}},\ }\href {\doibase
  10.1103/PhysRevLett.109.067205} {\bibfield  {journal} {\bibinfo  {journal}
  {Phys. Rev. Lett.}\ }\textbf {\bibinfo {volume} {109}},\ \bibinfo {pages}
  {067205} (\bibinfo {year} {2012})}\BibitemShut {NoStop}%
\bibitem [{\citenamefont {Bai}\ \emph {et~al.}(2005)\citenamefont {Bai},
  \citenamefont {Wang}, \citenamefont {Wuttig}, \citenamefont {Li},
  \citenamefont {Wang}, \citenamefont {Pyatakov}, \citenamefont {Zvezdin},
  \citenamefont {Cross},\ and\ \citenamefont {Viehland}}]{Bai2005}%
  \BibitemOpen
  \bibfield  {author} {\bibinfo {author} {\bibfnamefont {F.}~\bibnamefont
  {Bai}}, \bibinfo {author} {\bibfnamefont {J.}~\bibnamefont {Wang}}, \bibinfo
  {author} {\bibfnamefont {M.}~\bibnamefont {Wuttig}}, \bibinfo {author}
  {\bibfnamefont {J.}~\bibnamefont {Li}}, \bibinfo {author} {\bibfnamefont
  {N.}~\bibnamefont {Wang}}, \bibinfo {author} {\bibfnamefont {A.~P.}\
  \bibnamefont {Pyatakov}}, \bibinfo {author} {\bibfnamefont {A.~K.}\
  \bibnamefont {Zvezdin}}, \bibinfo {author} {\bibfnamefont {L.~E.}\
  \bibnamefont {Cross}}, \ and\ \bibinfo {author} {\bibfnamefont
  {D.}~\bibnamefont {Viehland}},\ }\href {\doibase 10.1063/1.1851612}
  {\bibfield  {journal} {\bibinfo  {journal} {Applied Physics Letters}\
  }\textbf {\bibinfo {volume} {86}},\ \bibinfo {eid} {032511} (\bibinfo {year}
  {2005})}\BibitemShut {NoStop}%
\bibitem [{\citenamefont {Chen}\ \emph {et~al.}(2012)\citenamefont {Chen},
  \citenamefont {G\"unayd\ifmmode \imath \else \i
  \fi{}n-\ifmmode~\mbox{\c{S}}\else \c{S}\fi{}en}, \citenamefont {Ren},
  \citenamefont {Qin}, \citenamefont {Brinzari}, \citenamefont {McGill},
  \citenamefont {Cheong},\ and\ \citenamefont {Musfeldt}}]{Chen2012}%
  \BibitemOpen
  \bibfield  {author} {\bibinfo {author} {\bibfnamefont {P.}~\bibnamefont
  {Chen}}, \bibinfo {author} {\bibfnamefont {O.}~\bibnamefont {G\"unayd\ifmmode
  \imath \else \i \fi{}n-\ifmmode~\mbox{\c{S}}\else \c{S}\fi{}en}}, \bibinfo
  {author} {\bibfnamefont {W.~J.}\ \bibnamefont {Ren}}, \bibinfo {author}
  {\bibfnamefont {Z.}~\bibnamefont {Qin}}, \bibinfo {author} {\bibfnamefont
  {T.~V.}\ \bibnamefont {Brinzari}}, \bibinfo {author} {\bibfnamefont
  {S.}~\bibnamefont {McGill}}, \bibinfo {author} {\bibfnamefont {S.-W.}\
  \bibnamefont {Cheong}}, \ and\ \bibinfo {author} {\bibfnamefont {J.~L.}\
  \bibnamefont {Musfeldt}},\ }\href {\doibase 10.1103/PhysRevB.86.014407}
  {\bibfield  {journal} {\bibinfo  {journal} {Phys. Rev. B}\ }\textbf {\bibinfo
  {volume} {86}},\ \bibinfo {pages} {014407} (\bibinfo {year}
  {2012})}\BibitemShut {NoStop}%
\bibitem [{\citenamefont {Tokunaga}\ \emph {et~al.}(2010)\citenamefont
  {Tokunaga}, \citenamefont {Azuma},\ and\ \citenamefont
  {Shimakawa}}]{Tokunaga2010JPSJ}%
  \BibitemOpen
  \bibfield  {author} {\bibinfo {author} {\bibfnamefont {M.}~\bibnamefont
  {Tokunaga}}, \bibinfo {author} {\bibfnamefont {M.}~\bibnamefont {Azuma}}, \
  and\ \bibinfo {author} {\bibfnamefont {Y.}~\bibnamefont {Shimakawa}},\ }\href
  {\doibase 10.1143/JPSJ.79.064713} {\bibfield  {journal} {\bibinfo  {journal}
  {J. Phys. Soc. Jpn.}\ }\textbf {\bibinfo {volume} {79}},\ \bibinfo {pages}
  {064713} (\bibinfo {year} {2010})}\BibitemShut {NoStop}%
\bibitem [{\citenamefont {Park}\ \emph {et~al.}(2011)\citenamefont {Park},
  \citenamefont {Lee}, \citenamefont {Lee}, \citenamefont {Gozzo},
  \citenamefont {Kimura}, \citenamefont {Noda}, \citenamefont {Choi},
  \citenamefont {Kiryukhin}, \citenamefont {Cheong}, \citenamefont {Jo},
  \citenamefont {Choi}, \citenamefont {Balicas}, \citenamefont {Jeon},\ and\
  \citenamefont {Park}}]{Park2011}%
  \BibitemOpen
  \bibfield  {author} {\bibinfo {author} {\bibfnamefont {J.}~\bibnamefont
  {Park}}, \bibinfo {author} {\bibfnamefont {S.-H.}\ \bibnamefont {Lee}},
  \bibinfo {author} {\bibfnamefont {S.}~\bibnamefont {Lee}}, \bibinfo {author}
  {\bibfnamefont {F.}~\bibnamefont {Gozzo}}, \bibinfo {author} {\bibfnamefont
  {H.}~\bibnamefont {Kimura}}, \bibinfo {author} {\bibfnamefont
  {Y.}~\bibnamefont {Noda}}, \bibinfo {author} {\bibfnamefont {Y.~J.}\
  \bibnamefont {Choi}}, \bibinfo {author} {\bibfnamefont {V.}~\bibnamefont
  {Kiryukhin}}, \bibinfo {author} {\bibfnamefont {S.-W.}\ \bibnamefont
  {Cheong}}, \bibinfo {author} {\bibfnamefont {Y.}~\bibnamefont {Jo}}, \bibinfo
  {author} {\bibfnamefont {E.~S.}\ \bibnamefont {Choi}}, \bibinfo {author}
  {\bibfnamefont {L.}~\bibnamefont {Balicas}}, \bibinfo {author} {\bibfnamefont
  {G.~S.}\ \bibnamefont {Jeon}}, \ and\ \bibinfo {author} {\bibfnamefont
  {J.-G.}\ \bibnamefont {Park}},\ }\href {\doibase 10.1143/JPSJ.80.114714}
  {\bibfield  {journal} {\bibinfo  {journal} {J. Phys. Soc. Jpn.}\ }\textbf
  {\bibinfo {volume} {80}},\ \bibinfo {pages} {114714} (\bibinfo {year}
  {2011})}\BibitemShut {NoStop}%
\bibitem [{\citenamefont {Fishman}\ \emph {et~al.}(2012)\citenamefont
  {Fishman}, \citenamefont {Furukawa}, \citenamefont {Haraldsen}, \citenamefont
  {Matsuda},\ and\ \citenamefont {Miyahara}}]{Fishman2012}%
  \BibitemOpen
  \bibfield  {author} {\bibinfo {author} {\bibfnamefont {R.~S.}\ \bibnamefont
  {Fishman}}, \bibinfo {author} {\bibfnamefont {N.}~\bibnamefont {Furukawa}},
  \bibinfo {author} {\bibfnamefont {J.~T.}\ \bibnamefont {Haraldsen}}, \bibinfo
  {author} {\bibfnamefont {M.}~\bibnamefont {Matsuda}}, \ and\ \bibinfo
  {author} {\bibfnamefont {S.}~\bibnamefont {Miyahara}},\ }\href {\doibase
  10.1103/PhysRevB.86.220402} {\bibfield  {journal} {\bibinfo  {journal} {Phys.
  Rev. B}\ }\textbf {\bibinfo {volume} {86}},\ \bibinfo {pages} {220402}
  (\bibinfo {year} {2012})}\BibitemShut {NoStop}%
\bibitem [{\citenamefont {Talbayev}\ \emph {et~al.}(2011)\citenamefont
  {Talbayev}, \citenamefont {Trugman}, \citenamefont {Lee}, \citenamefont {Yi},
  \citenamefont {Cheong},\ and\ \citenamefont {Taylor}}]{Talbayev2011}%
  \BibitemOpen
  \bibfield  {author} {\bibinfo {author} {\bibfnamefont {D.}~\bibnamefont
  {Talbayev}}, \bibinfo {author} {\bibfnamefont {S.~A.}\ \bibnamefont
  {Trugman}}, \bibinfo {author} {\bibfnamefont {S.}~\bibnamefont {Lee}},
  \bibinfo {author} {\bibfnamefont {H.~T.}\ \bibnamefont {Yi}}, \bibinfo
  {author} {\bibfnamefont {S.-W.}\ \bibnamefont {Cheong}}, \ and\ \bibinfo
  {author} {\bibfnamefont {A.~J.}\ \bibnamefont {Taylor}},\ }\href {\doibase
  10.1103/PhysRevB.83.094403} {\bibfield  {journal} {\bibinfo  {journal} {Phys.
  Rev. B}\ }\textbf {\bibinfo {volume} {83}},\ \bibinfo {pages} {094403}
  (\bibinfo {year} {2011})}\BibitemShut {NoStop}%
\bibitem [{\citenamefont {H\"{u}vonen}\ \emph {et~al.}(2009)\citenamefont
  {H\"{u}vonen}, \citenamefont {Nagel}, \citenamefont {R{\~o}{\~o}m},
  \citenamefont {Choi}, \citenamefont {Zhang}, \citenamefont {Park},\ and\
  \citenamefont {Cheong}}]{Huvonen2009}%
  \BibitemOpen
  \bibfield  {author} {\bibinfo {author} {\bibfnamefont {D.}~\bibnamefont
  {H\"{u}vonen}}, \bibinfo {author} {\bibfnamefont {U.}~\bibnamefont {Nagel}},
  \bibinfo {author} {\bibfnamefont {T.}~\bibnamefont {R{\~o}{\~o}m}}, \bibinfo
  {author} {\bibfnamefont {Y.~J.}\ \bibnamefont {Choi}}, \bibinfo {author}
  {\bibfnamefont {C.~L.}\ \bibnamefont {Zhang}}, \bibinfo {author}
  {\bibfnamefont {S.}~\bibnamefont {Park}}, \ and\ \bibinfo {author}
  {\bibfnamefont {S.-W.}\ \bibnamefont {Cheong}},\ }\href {\doibase
  10.1103/PhysRevB.80.100402} {\bibfield  {journal} {\bibinfo  {journal} {Phys.
  Rev. B}\ }\textbf {\bibinfo {volume} {80}},\ \bibinfo {eid} {100402}
  (\bibinfo {year} {2009})}\BibitemShut {NoStop}%
\bibitem [{\citenamefont {Kadomtseva}\ \emph {et~al.}(2004)\citenamefont
  {Kadomtseva}, \citenamefont {Zvezdin}, \citenamefont {Popov}, \citenamefont
  {Pyatakov},\ and\ \citenamefont {Vorobev}}]{Kadomtseva2004}%
  \BibitemOpen
  \bibfield  {author} {\bibinfo {author} {\bibfnamefont {A.~M.}\ \bibnamefont
  {Kadomtseva}}, \bibinfo {author} {\bibfnamefont {A.~K.}\ \bibnamefont
  {Zvezdin}}, \bibinfo {author} {\bibfnamefont {Y.~P.}\ \bibnamefont {Popov}},
  \bibinfo {author} {\bibfnamefont {A.~P.}\ \bibnamefont {Pyatakov}}, \ and\
  \bibinfo {author} {\bibfnamefont {G.~P.}\ \bibnamefont {Vorobev}},\
  }\href@noop {} {\bibfield  {journal} {\bibinfo  {journal} {JETP Lett.}\
  }\textbf {\bibinfo {volume} {79}},\ \bibinfo {pages} {571} (\bibinfo {year}
  {2004})}\BibitemShut {NoStop}%
\bibitem [{\citenamefont {Ederer}\ and\ \citenamefont
  {Spaldin}(2005)}]{Ederer2005}%
  \BibitemOpen
  \bibfield  {author} {\bibinfo {author} {\bibfnamefont {C.}~\bibnamefont
  {Ederer}}\ and\ \bibinfo {author} {\bibfnamefont {N.~A.}\ \bibnamefont
  {Spaldin}},\ }\href {\doibase 10.1103/PhysRevB.71.060401} {\bibfield
  {journal} {\bibinfo  {journal} {Phys. Rev. B}\ }\textbf {\bibinfo {volume}
  {71}},\ \bibinfo {pages} {060401} (\bibinfo {year} {2005})}\BibitemShut
  {NoStop}%
\bibitem [{\citenamefont {Albrecht}\ \emph {et~al.}(2010)\citenamefont
  {Albrecht}, \citenamefont {Lisenkov}, \citenamefont {Ren}, \citenamefont
  {Rahmedov}, \citenamefont {Kornev},\ and\ \citenamefont
  {Bellaiche}}]{Albrecht2010}%
  \BibitemOpen
  \bibfield  {author} {\bibinfo {author} {\bibfnamefont {D.}~\bibnamefont
  {Albrecht}}, \bibinfo {author} {\bibfnamefont {S.}~\bibnamefont {Lisenkov}},
  \bibinfo {author} {\bibfnamefont {W.}~\bibnamefont {Ren}}, \bibinfo {author}
  {\bibfnamefont {D.}~\bibnamefont {Rahmedov}}, \bibinfo {author}
  {\bibfnamefont {I.~A.}\ \bibnamefont {Kornev}}, \ and\ \bibinfo {author}
  {\bibfnamefont {L.}~\bibnamefont {Bellaiche}},\ }\href {\doibase
  10.1103/PhysRevB.81.140401} {\bibfield  {journal} {\bibinfo  {journal} {Phys.
  Rev. B}\ }\textbf {\bibinfo {volume} {81}},\ \bibinfo {pages} {140401}
  (\bibinfo {year} {2010})}\BibitemShut {NoStop}%
\bibitem [{\citenamefont {Ramazanoglu}\ \emph
  {et~al.}(2011{\natexlab{b}})\citenamefont {Ramazanoglu}, \citenamefont
  {Laver}, \citenamefont {Ratcliff}, \citenamefont {Watson}, \citenamefont
  {Chen}, \citenamefont {Jackson}, \citenamefont {Kothapalli}, \citenamefont
  {Lee}, \citenamefont {Cheong},\ and\ \citenamefont
  {Kiryukhin}}]{Ramazanoglu2011prl}%
  \BibitemOpen
  \bibfield  {author} {\bibinfo {author} {\bibfnamefont {M.}~\bibnamefont
  {Ramazanoglu}}, \bibinfo {author} {\bibfnamefont {M.}~\bibnamefont {Laver}},
  \bibinfo {author} {\bibfnamefont {W.}~\bibnamefont {Ratcliff}}, \bibinfo
  {author} {\bibfnamefont {S.~M.}\ \bibnamefont {Watson}}, \bibinfo {author}
  {\bibfnamefont {W.~C.}\ \bibnamefont {Chen}}, \bibinfo {author}
  {\bibfnamefont {A.}~\bibnamefont {Jackson}}, \bibinfo {author} {\bibfnamefont
  {K.}~\bibnamefont {Kothapalli}}, \bibinfo {author} {\bibfnamefont
  {S.}~\bibnamefont {Lee}}, \bibinfo {author} {\bibfnamefont {S.-W.}\
  \bibnamefont {Cheong}}, \ and\ \bibinfo {author} {\bibfnamefont
  {V.}~\bibnamefont {Kiryukhin}},\ }\href {\doibase
  10.1103/PhysRevLett.107.207206} {\bibfield  {journal} {\bibinfo  {journal}
  {Phys. Rev. Lett.}\ }\textbf {\bibinfo {volume} {107}},\ \bibinfo {pages}
  {207206} (\bibinfo {year} {2011}{\natexlab{b}})}\BibitemShut {NoStop}%
\bibitem [{\citenamefont {Przenios{\l}o}\ \emph {et~al.}(2006)\citenamefont
  {Przenios{\l}o}, \citenamefont {Palewicz}, \citenamefont {Regulski},
  \citenamefont {Sosnowska}, \citenamefont {Ibberson},\ and\ \citenamefont
  {Knight}}]{Przenioslo2006}%
  \BibitemOpen
  \bibfield  {author} {\bibinfo {author} {\bibfnamefont {R.}~\bibnamefont
  {Przenios{\l}o}}, \bibinfo {author} {\bibfnamefont {A.}~\bibnamefont
  {Palewicz}}, \bibinfo {author} {\bibfnamefont {M.}~\bibnamefont {Regulski}},
  \bibinfo {author} {\bibfnamefont {I.}~\bibnamefont {Sosnowska}}, \bibinfo
  {author} {\bibfnamefont {R.~M.}\ \bibnamefont {Ibberson}}, \ and\ \bibinfo
  {author} {\bibfnamefont {K.~S.}\ \bibnamefont {Knight}},\ }\href
  {http://stacks.iop.org/0953-8984/18/i=6/a=019} {\bibfield  {journal}
  {\bibinfo  {journal} {J. Phys.: Condens. Matter}\ }\textbf {\bibinfo {volume}
  {18}},\ \bibinfo {pages} {2069} (\bibinfo {year} {2006})}\BibitemShut
  {NoStop}%
\bibitem [{\citenamefont {Herrero-Albillos}\ \emph
  {et~al.}(2010{\natexlab{b}})\citenamefont {Herrero-Albillos}, \citenamefont
  {Catalan}, \citenamefont {Rodriguez-Velamazan}, \citenamefont {Viret},
  \citenamefont {Colson},\ and\ \citenamefont {Scott}}]{Albillos2010}%
  \BibitemOpen
  \bibfield  {author} {\bibinfo {author} {\bibfnamefont {J.}~\bibnamefont
  {Herrero-Albillos}}, \bibinfo {author} {\bibfnamefont {G.}~\bibnamefont
  {Catalan}}, \bibinfo {author} {\bibfnamefont {J.~A.}\ \bibnamefont
  {Rodriguez-Velamazan}}, \bibinfo {author} {\bibfnamefont {M.}~\bibnamefont
  {Viret}}, \bibinfo {author} {\bibfnamefont {D.}~\bibnamefont {Colson}}, \
  and\ \bibinfo {author} {\bibfnamefont {J.~F.}\ \bibnamefont {Scott}},\ }\href
  {http://stacks.iop.org/0953-8984/22/i=25/a=256001} {\bibfield  {journal}
  {\bibinfo  {journal} {J. Phys.: Condens. Matter}\ }\textbf {\bibinfo {volume}
  {22}},\ \bibinfo {pages} {256001} (\bibinfo {year}
  {2010}{\natexlab{b}})}\BibitemShut {NoStop}%
\bibitem [{\citenamefont {Kadomtseva}\ \emph {et~al.}(2006)\citenamefont
  {Kadomtseva}, \citenamefont {Popov}, \citenamefont {Pyatakov}, \citenamefont
  {Vorob'ev}, \citenamefont {Zvezdin},\ and\ \citenamefont
  {Viehland}}]{Kadomtseva2006}%
  \BibitemOpen
  \bibfield  {author} {\bibinfo {author} {\bibfnamefont {A.~M.}\ \bibnamefont
  {Kadomtseva}}, \bibinfo {author} {\bibfnamefont {Y.~P.}\ \bibnamefont
  {Popov}}, \bibinfo {author} {\bibfnamefont {A.~P.}\ \bibnamefont {Pyatakov}},
  \bibinfo {author} {\bibfnamefont {G.~P.}\ \bibnamefont {Vorob'ev}}, \bibinfo
  {author} {\bibfnamefont {A.~K.}\ \bibnamefont {Zvezdin}}, \ and\ \bibinfo
  {author} {\bibfnamefont {D.}~\bibnamefont {Viehland}},\ }\href@noop {}
  {\bibfield  {journal} {\bibinfo  {journal} {Phase Transitions}\ }\textbf
  {\bibinfo {volume} {79}},\ \bibinfo {pages} {1019} (\bibinfo {year}
  {2006})}\BibitemShut {NoStop}%
\bibitem [{\citenamefont {Tehranchi}\ \emph {et~al.}(1997)\citenamefont
  {Tehranchi}, \citenamefont {Kubrakov},\ and\ \citenamefont
  {Zvezdin}}]{Tehranchi1997}%
  \BibitemOpen
  \bibfield  {author} {\bibinfo {author} {\bibfnamefont {M.-M.}\ \bibnamefont
  {Tehranchi}}, \bibinfo {author} {\bibfnamefont {N.~F.}\ \bibnamefont
  {Kubrakov}}, \ and\ \bibinfo {author} {\bibfnamefont {A.~K.}\ \bibnamefont
  {Zvezdin}},\ }\href@noop {} {\bibfield  {journal} {\bibinfo  {journal}
  {Ferroelectrics}\ }\textbf {\bibinfo {volume} {204}},\ \bibinfo {pages} {181}
  (\bibinfo {year} {1997})}\BibitemShut {NoStop}%
\bibitem [{\citenamefont {Ohoyama}\ \emph {et~al.}(2011)\citenamefont
  {Ohoyama}, \citenamefont {Lee}, \citenamefont {Yoshii}, \citenamefont
  {Narumi}, \citenamefont {Morioka}, \citenamefont {Nojiri}, \citenamefont
  {Jeon}, \citenamefont {Cheong},\ and\ \citenamefont {Park}}]{Ohoyama2011}%
  \BibitemOpen
  \bibfield  {author} {\bibinfo {author} {\bibfnamefont {K.}~\bibnamefont
  {Ohoyama}}, \bibinfo {author} {\bibfnamefont {S.}~\bibnamefont {Lee}},
  \bibinfo {author} {\bibfnamefont {S.}~\bibnamefont {Yoshii}}, \bibinfo
  {author} {\bibfnamefont {Y.}~\bibnamefont {Narumi}}, \bibinfo {author}
  {\bibfnamefont {T.}~\bibnamefont {Morioka}}, \bibinfo {author} {\bibfnamefont
  {H.}~\bibnamefont {Nojiri}}, \bibinfo {author} {\bibfnamefont {G.~S.}\
  \bibnamefont {Jeon}}, \bibinfo {author} {\bibfnamefont {S.-W.}\ \bibnamefont
  {Cheong}}, \ and\ \bibinfo {author} {\bibfnamefont {J.-G.}\ \bibnamefont
  {Park}},\ }\href {\doibase 10.1143/JPSJ.80.125001} {\bibfield  {journal}
  {\bibinfo  {journal} {J. Phys. Soc. Jpn.}\ }\textbf {\bibinfo {volume}
  {80}},\ \bibinfo {pages} {125001} (\bibinfo {year} {2011})}\BibitemShut
  {NoStop}%
\bibitem [{\citenamefont {de~Sousa}\ and\ \citenamefont
  {Moore}(2008)}]{DeSousa2008}%
  \BibitemOpen
  \bibfield  {author} {\bibinfo {author} {\bibfnamefont {R.}~\bibnamefont
  {de~Sousa}}\ and\ \bibinfo {author} {\bibfnamefont {J.~E.}\ \bibnamefont
  {Moore}},\ }\href {\doibase 10.1103/PhysRevB.77.012406} {\bibfield  {journal}
  {\bibinfo  {journal} {Phys. Rev. B}\ }\textbf {\bibinfo {volume} {77}},\
  \bibinfo {pages} {012406} (\bibinfo {year} {2008})}\BibitemShut {NoStop}%
\bibitem [{\citenamefont {Fishman}\ \emph {et~al.}(2013)\citenamefont
  {Fishman}, \citenamefont {Haraldsen}, \citenamefont {Furukawa},\ and\
  \citenamefont {Miyahara}}]{Fishman2013PRB}%
  \BibitemOpen
  \bibfield  {author} {\bibinfo {author} {\bibfnamefont {R.~S.}\ \bibnamefont
  {Fishman}}, \bibinfo {author} {\bibfnamefont {J.~T.}\ \bibnamefont
  {Haraldsen}}, \bibinfo {author} {\bibfnamefont {N.}~\bibnamefont {Furukawa}},
  \ and\ \bibinfo {author} {\bibfnamefont {S.}~\bibnamefont {Miyahara}},\
  }\href {\doibase 10.1103/PhysRevB.87.134416} {\bibfield  {journal} {\bibinfo
  {journal} {Phys. Rev. B}\ }\textbf {\bibinfo {volume} {87}},\ \bibinfo
  {pages} {134416} (\bibinfo {year} {2013})}\BibitemShut {NoStop}%
\bibitem [{\citenamefont {Rovillain}\ \emph {et~al.}(2010)\citenamefont
  {Rovillain}, \citenamefont {Sousa}, \citenamefont {Gallais}, \citenamefont
  {Sacuto}, \citenamefont {M{\'e}asson}, \citenamefont {Colson}, \citenamefont
  {Forget}, \citenamefont {M.~Bibes}, \citenamefont {Barth{\'e}l{\'e}my},\ and\
  \citenamefont {Cazayous}}]{Rovillain2010}%
  \BibitemOpen
  \bibfield  {author} {\bibinfo {author} {\bibfnamefont {P.}~\bibnamefont
  {Rovillain}}, \bibinfo {author} {\bibfnamefont {R.~d.}\ \bibnamefont
  {Sousa}}, \bibinfo {author} {\bibfnamefont {Y.}~\bibnamefont {Gallais}},
  \bibinfo {author} {\bibfnamefont {A.}~\bibnamefont {Sacuto}}, \bibinfo
  {author} {\bibfnamefont {M.~A.}\ \bibnamefont {M{\'e}asson}}, \bibinfo
  {author} {\bibfnamefont {D.}~\bibnamefont {Colson}}, \bibinfo {author}
  {\bibfnamefont {A.}~\bibnamefont {Forget}}, \bibinfo {author} {\bibfnamefont
  {M.}~\bibnamefont {M.~Bibes}}, \bibinfo {author} {\bibfnamefont
  {A.}~\bibnamefont {Barth{\'e}l{\'e}my}}, \ and\ \bibinfo {author}
  {\bibfnamefont {M.}~\bibnamefont {Cazayous}},\ }\href@noop {} {\bibfield
  {journal} {\bibinfo  {journal} {Nature Mater.}\ }\textbf {\bibinfo {volume}
  {9}},\ \bibinfo {pages} {975} (\bibinfo {year} {2010})}\BibitemShut {NoStop}%
\bibitem [{\citenamefont {Cazayous}\ \emph {et~al.}(2008)\citenamefont
  {Cazayous}, \citenamefont {Gallais}, \citenamefont {Sacuto}, \citenamefont
  {de~Sousa}, \citenamefont {Lebeugle},\ and\ \citenamefont
  {Colson}}]{Cazayous2008}%
  \BibitemOpen
  \bibfield  {author} {\bibinfo {author} {\bibfnamefont {M.}~\bibnamefont
  {Cazayous}}, \bibinfo {author} {\bibfnamefont {Y.}~\bibnamefont {Gallais}},
  \bibinfo {author} {\bibfnamefont {A.}~\bibnamefont {Sacuto}}, \bibinfo
  {author} {\bibfnamefont {R.}~\bibnamefont {de~Sousa}}, \bibinfo {author}
  {\bibfnamefont {D.}~\bibnamefont {Lebeugle}}, \ and\ \bibinfo {author}
  {\bibfnamefont {D.}~\bibnamefont {Colson}},\ }\href {\doibase
  10.1103/PhysRevLett.101.037601} {\bibfield  {journal} {\bibinfo  {journal}
  {Phys. Rev. Lett.}\ }\textbf {\bibinfo {volume} {101}},\ \bibinfo {pages}
  {037601} (\bibinfo {year} {2008})}\BibitemShut {NoStop}%
\bibitem [{\citenamefont {Rovillain}\ \emph {et~al.}(2009)\citenamefont
  {Rovillain}, \citenamefont {Cazayous}, \citenamefont {Gallais}, \citenamefont
  {Sacuto}, \citenamefont {Lobo}, \citenamefont {Lebeugle},\ and\ \citenamefont
  {Colson}}]{Rovillain2009}%
  \BibitemOpen
  \bibfield  {author} {\bibinfo {author} {\bibfnamefont {P.}~\bibnamefont
  {Rovillain}}, \bibinfo {author} {\bibfnamefont {M.}~\bibnamefont {Cazayous}},
  \bibinfo {author} {\bibfnamefont {Y.}~\bibnamefont {Gallais}}, \bibinfo
  {author} {\bibfnamefont {A.}~\bibnamefont {Sacuto}}, \bibinfo {author}
  {\bibfnamefont {R.~P. S.~M.}\ \bibnamefont {Lobo}}, \bibinfo {author}
  {\bibfnamefont {D.}~\bibnamefont {Lebeugle}}, \ and\ \bibinfo {author}
  {\bibfnamefont {D.}~\bibnamefont {Colson}},\ }\href {\doibase
  10.1103/PhysRevB.79.180411} {\bibfield  {journal} {\bibinfo  {journal} {Phys.
  Rev. B}\ }\textbf {\bibinfo {volume} {79}},\ \bibinfo {eid} {180411}
  (\bibinfo {year} {2009})}\BibitemShut {NoStop}%
\bibitem [{\citenamefont {Komandin}\ \emph {et~al.}(2010)\citenamefont
  {Komandin}, \citenamefont {Torgashev}, \citenamefont {Volkov}, \citenamefont
  {O.~E.~Porodinkov}, \citenamefont {Spektor},\ and\ \citenamefont
  {Bush}}]{Komandin2010}%
  \BibitemOpen
  \bibfield  {author} {\bibinfo {author} {\bibfnamefont {G.~A.}\ \bibnamefont
  {Komandin}}, \bibinfo {author} {\bibfnamefont {V.~I.}\ \bibnamefont
  {Torgashev}}, \bibinfo {author} {\bibfnamefont {A.~A.}\ \bibnamefont
  {Volkov}}, \bibinfo {author} {\bibfnamefont {O.~E.}\ \bibnamefont
  {O.~E.~Porodinkov}}, \bibinfo {author} {\bibfnamefont {I.~E.}\ \bibnamefont
  {Spektor}}, \ and\ \bibinfo {author} {\bibfnamefont {A.~A.}\ \bibnamefont
  {Bush}},\ }\href@noop {} {\bibfield  {journal} {\bibinfo  {journal} {Physics
  of the Solid State}\ }\textbf {\bibinfo {volume} {52}},\ \bibinfo {pages}
  {734} (\bibinfo {year} {2010})}\BibitemShut {NoStop}%
\bibitem [{\citenamefont {Choi}\ \emph {et~al.}(2009)\citenamefont {Choi},
  \citenamefont {Lee}, \citenamefont {Choi}, \citenamefont {Kiryukhin},\ and\
  \citenamefont {Cheong}}]{Choi2009}%
  \BibitemOpen
  \bibfield  {author} {\bibinfo {author} {\bibfnamefont {T.}~\bibnamefont
  {Choi}}, \bibinfo {author} {\bibfnamefont {S.}~\bibnamefont {Lee}}, \bibinfo
  {author} {\bibfnamefont {Y.~J.}\ \bibnamefont {Choi}}, \bibinfo {author}
  {\bibfnamefont {V.}~\bibnamefont {Kiryukhin}}, \ and\ \bibinfo {author}
  {\bibfnamefont {S.-W.}\ \bibnamefont {Cheong}},\ }\href@noop {} {\bibfield
  {journal} {\bibinfo  {journal} {Science}\ }\textbf {\bibinfo {volume}
  {324}},\ \bibinfo {pages} {63} (\bibinfo {year} {2009})}\BibitemShut
  {NoStop}%
\bibitem [{\citenamefont {Talbayev}\ \emph {et~al.}(2008)\citenamefont
  {Talbayev}, \citenamefont {LaForge}, \citenamefont {Trugman}, \citenamefont
  {Hur}, \citenamefont {Taylor}, \citenamefont {Averitt},\ and\ \citenamefont
  {Basov}}]{Talbayev2008}%
  \BibitemOpen
  \bibfield  {author} {\bibinfo {author} {\bibfnamefont {D.}~\bibnamefont
  {Talbayev}}, \bibinfo {author} {\bibfnamefont {A.~D.}\ \bibnamefont
  {LaForge}}, \bibinfo {author} {\bibfnamefont {S.~A.}\ \bibnamefont
  {Trugman}}, \bibinfo {author} {\bibfnamefont {N.}~\bibnamefont {Hur}},
  \bibinfo {author} {\bibfnamefont {A.~J.}\ \bibnamefont {Taylor}}, \bibinfo
  {author} {\bibfnamefont {R.~D.}\ \bibnamefont {Averitt}}, \ and\ \bibinfo
  {author} {\bibfnamefont {D.~N.}\ \bibnamefont {Basov}},\ }\href {\doibase
  10.1103/PhysRevLett.101.247601} {\bibfield  {journal} {\bibinfo  {journal}
  {Phys. Rev. Lett.}\ }\textbf {\bibinfo {volume} {101}},\ \bibinfo {eid}
  {247601} (\bibinfo {year} {2008})}\BibitemShut {NoStop}%
\bibitem [{\citenamefont {R{\~o}{\~o}m}\ \emph {et~al.}(2004)\citenamefont
  {R{\~o}{\~o}m}, \citenamefont {H{\"u}vonen}, \citenamefont {Nagel},
  \citenamefont {Wang},\ and\ \citenamefont {Kremer}}]{room2004NaVa}%
  \BibitemOpen
  \bibfield  {author} {\bibinfo {author} {\bibfnamefont {T.}~\bibnamefont
  {R{\~o}{\~o}m}}, \bibinfo {author} {\bibfnamefont {D.}~\bibnamefont
  {H{\"u}vonen}}, \bibinfo {author} {\bibfnamefont {U.}~\bibnamefont {Nagel}},
  \bibinfo {author} {\bibfnamefont {Y.-J.}\ \bibnamefont {Wang}}, \ and\
  \bibinfo {author} {\bibfnamefont {R.~K.}\ \bibnamefont {Kremer}},\ }\href
  {http://link.aps.org/abstract/PRB/v69/e144410} {\bibfield  {journal}
  {\bibinfo  {journal} {Phys. Rev. B}\ }\textbf {\bibinfo {volume} {69}},\
  \bibinfo {eid} {144410} (\bibinfo {year} {2004})}\BibitemShut {NoStop}%
\bibitem [{PRL()}]{PRL2013suppl}%
  \BibitemOpen
  \href@noop {} {}\bibinfo {note} {Supplementary online
  information}\BibitemShut {NoStop}%
\bibitem [{\citenamefont {Ruette}\ \emph {et~al.}(2004)\citenamefont {Ruette},
  \citenamefont {Zvyagin}, \citenamefont {Pyatakov}, \citenamefont {Bush},
  \citenamefont {Li}, \citenamefont {Belotelov}, \citenamefont {Zvezdin},\ and\
  \citenamefont {Viehland}}]{Ruette2004}%
  \BibitemOpen
  \bibfield  {author} {\bibinfo {author} {\bibfnamefont {B.}~\bibnamefont
  {Ruette}}, \bibinfo {author} {\bibfnamefont {S.}~\bibnamefont {Zvyagin}},
  \bibinfo {author} {\bibfnamefont {A.~P.}\ \bibnamefont {Pyatakov}}, \bibinfo
  {author} {\bibfnamefont {A.}~\bibnamefont {Bush}}, \bibinfo {author}
  {\bibfnamefont {J.~F.}\ \bibnamefont {Li}}, \bibinfo {author} {\bibfnamefont
  {V.~I.}\ \bibnamefont {Belotelov}}, \bibinfo {author} {\bibfnamefont {A.~K.}\
  \bibnamefont {Zvezdin}}, \ and\ \bibinfo {author} {\bibfnamefont
  {D.}~\bibnamefont {Viehland}},\ }\href {\doibase 10.1103/PhysRevB.69.064114}
  {\bibfield  {journal} {\bibinfo  {journal} {Phys. Rev. B}\ }\textbf {\bibinfo
  {volume} {69}},\ \bibinfo {pages} {064114} (\bibinfo {year}
  {2004})}\BibitemShut {NoStop}%
\bibitem [{\citenamefont {Xu}\ \emph {et~al.}(2009)\citenamefont {Xu},
  \citenamefont {Brinzari}, \citenamefont {Lee}, \citenamefont {Chu},
  \citenamefont {Martin}, \citenamefont {Kumar}, \citenamefont {McGill},
  \citenamefont {Rai}, \citenamefont {Ramesh}, \citenamefont {Gopalan},
  \citenamefont {Cheong},\ and\ \citenamefont {Musfeldt}}]{Xu2009bfo}%
  \BibitemOpen
  \bibfield  {author} {\bibinfo {author} {\bibfnamefont {X.~S.}\ \bibnamefont
  {Xu}}, \bibinfo {author} {\bibfnamefont {T.~V.}\ \bibnamefont {Brinzari}},
  \bibinfo {author} {\bibfnamefont {S.}~\bibnamefont {Lee}}, \bibinfo {author}
  {\bibfnamefont {Y.~H.}\ \bibnamefont {Chu}}, \bibinfo {author} {\bibfnamefont
  {L.~W.}\ \bibnamefont {Martin}}, \bibinfo {author} {\bibfnamefont
  {A.}~\bibnamefont {Kumar}}, \bibinfo {author} {\bibfnamefont
  {S.}~\bibnamefont {McGill}}, \bibinfo {author} {\bibfnamefont {R.~C.}\
  \bibnamefont {Rai}}, \bibinfo {author} {\bibfnamefont {R.}~\bibnamefont
  {Ramesh}}, \bibinfo {author} {\bibfnamefont {V.}~\bibnamefont {Gopalan}},
  \bibinfo {author} {\bibfnamefont {S.~W.}\ \bibnamefont {Cheong}}, \ and\
  \bibinfo {author} {\bibfnamefont {J.~L.}\ \bibnamefont {Musfeldt}},\ }\href
  {\doibase 10.1103/PhysRevB.79.134425} {\bibfield  {journal} {\bibinfo
  {journal} {Phys. Rev. B}\ }\textbf {\bibinfo {volume} {79}},\ \bibinfo
  {pages} {134425} (\bibinfo {year} {2009})}\BibitemShut {NoStop}%
\end{thebibliography}

%

\section{Supplementary online information}

\begin{figure}[b]
%
\includegraphics[width=0.49\textwidth]{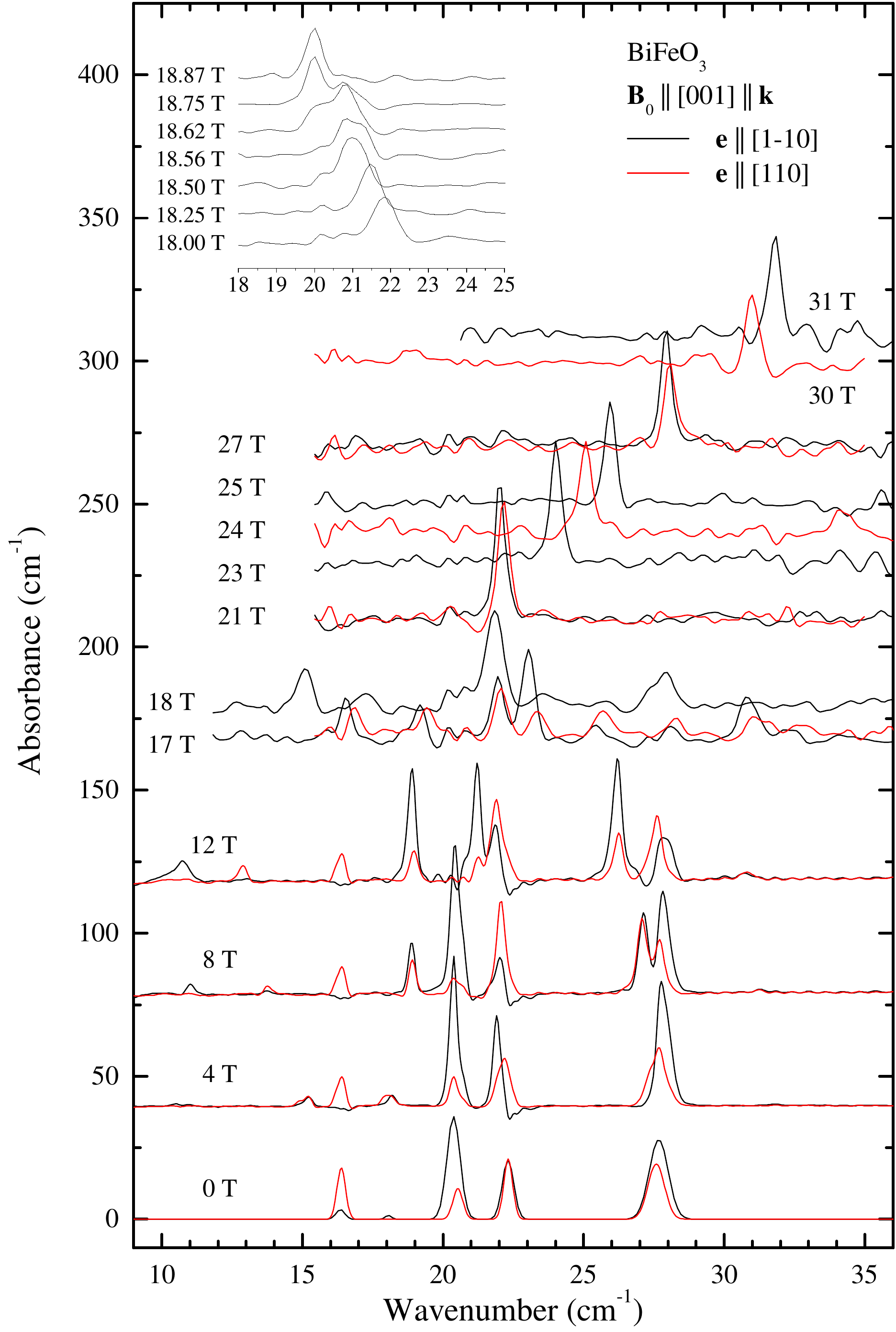}
\caption{\label{fig:BFOaFarFieldAbs} 
Magnetic field dependence of absorbance spectra of \BFO/ in $\mathbf{e}\parallel [1\bar{1}0]$ (black) and $\mathbf{e}\parallel [110]$ (red) polarizations.
Zero field spectra are fits as described in the main text. 
The inset shows a detailed view in the spectral range of the AFM resonance close to the metamagnetic transition, $B_c=18.8$\,T.
The spectra were measured after the sample had been in high field $\geq 12$\,T.
} 
\end{figure}

\begin{figure}
%
\includegraphics[width=0.49\textwidth]{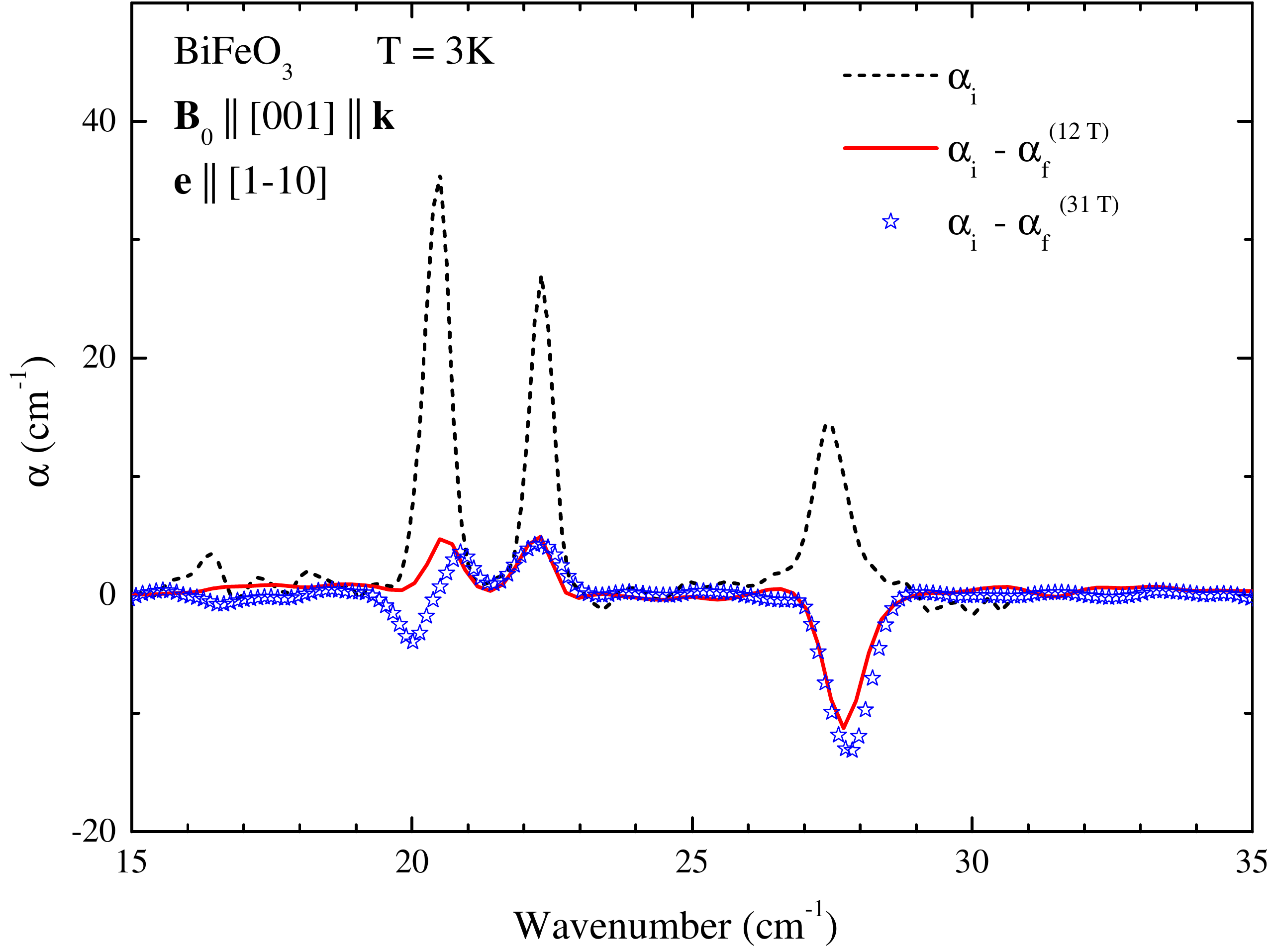}
\caption{\label{fig:zf12T31T}
This figure demonstrates how the application of magnetic field in [001] direction at low $T$ changes the zero field spectrum.
The dotted line is the zero field spectrum $\alpha_i$ measured after the sample was cooled in zero magnetic field.
We measured the zero field spectrum again after applying 12\,T at low temperature and observed a change in the spectrum.
The red solid line is the difference of two spectra $\alpha_i - \alpha_f^{(12T)} $ where $\alpha_f^{(12T)}$ is the zero field spectrum measured after application of 12\,T, experiment done in Tallinn.
In the Nijmegen experiment we applied a 31\,T field, well above the value of the metamagnetic transition at $\approx 19 $\,T, which had a similar effect on the spectrum of \BFO/, measured in zero field after application of magnetic field.
Blue stars show  differential absorption $\alpha_i - \alpha_f^{(31T)} $ where $\alpha_f^{(31T)}$ is the zero field spectrum measured  after applying 31\,T.
Negative difference at the 28\wn/ means that this line becomes stronger after the application of magnetic field. 
The lines at 20.5 and 22.5\wn/ loose some intensity after application of magnetic field.
Relative change of 20.5\wn/ line is the smallest of three lines and shows a small shift after 31\,T has been applied.
} 
\end{figure}

\begin{figure}
%
\includegraphics[width=0.49\textwidth]{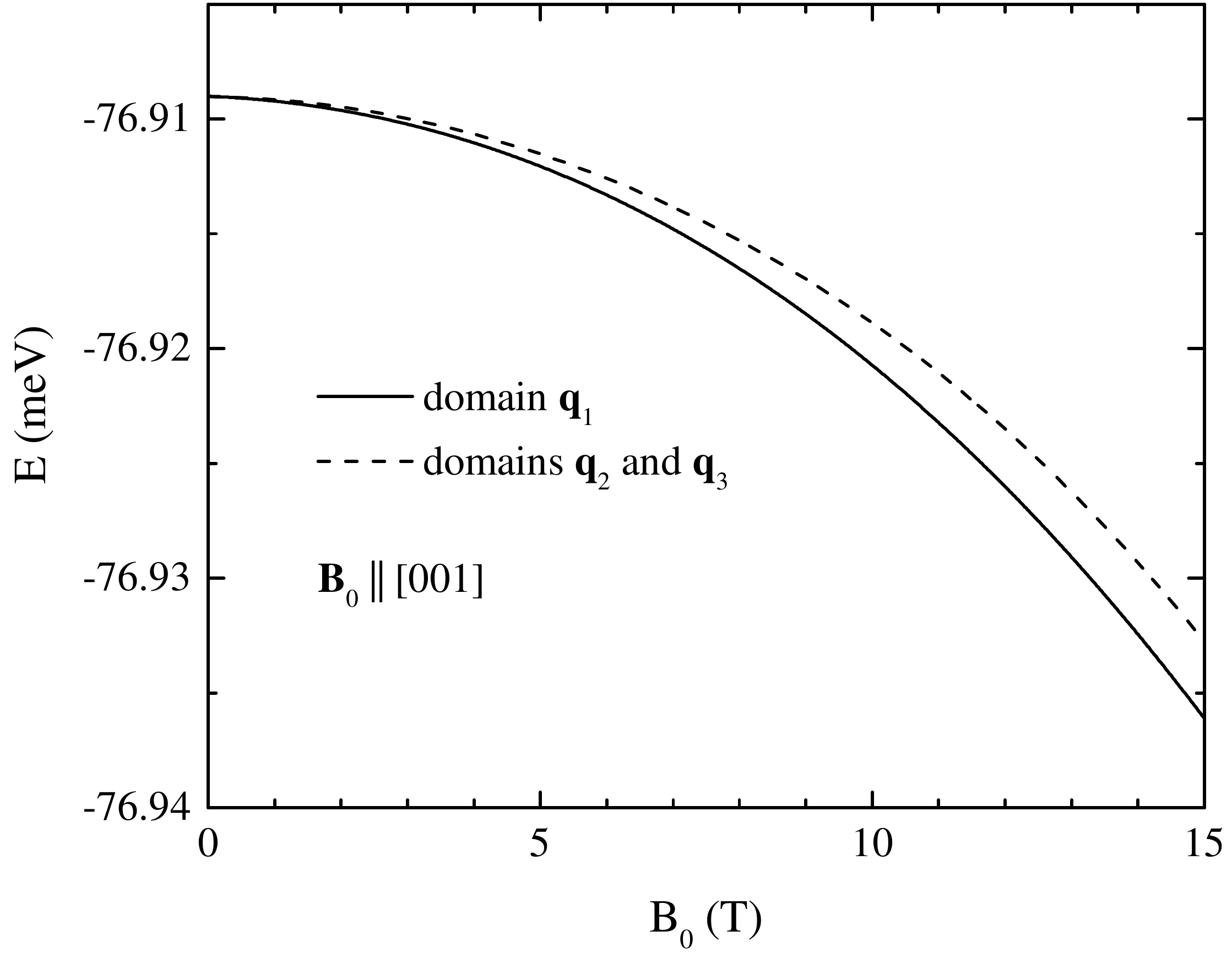}
\caption{\label{fig:Domain123Energy} 
This figure shows the calculated energy per site for cycloids $\mathbf{q}_1$, $\mathbf{q}_2$, and $\mathbf{q}_3$ in magnetic field along [001] direction.
} 
\end{figure}


\begin{table*}
\caption{\label{table:zero_field_areas}
Fitted positions and areas of absorption lines in zero magnetic field from Fig.\,2 of the main paper. 
$A_{i}$ and $A_f$ are the line areas before and after application of $B_0>21$\,T at low $T$.
Fit errors are three times the standard error.
The $\Phi_{1}^{(2)}$ mode position is an extrapolated value of the magnetic field dependence to 0\,T.
}
\begin{tabular}{|c|c|c|c|c|c|c|}
\hline 
 & \multicolumn{3}{c|}{} & \multicolumn{3}{c|}{} \\
 & \multicolumn{3}{c|}{$\mathbf{e}\,||\,[110]$} & \multicolumn{3}{c|}{$\mathbf{e}\,||\,[1\bar{1}0]$} \\
Mode & center / \wn/ & $A_f$ / \area/ & $A_{i}$ / \area/ 
& center / \wn/ & $A_f$ / \area/ & $A_{i}$ / \area/ \\ 
\hline

$\Phi_{1}^{(2)}$ & $11.0 \pm 0.1$ & 
- & - &-&-&-\\ 

$\Psi_0, \Phi_{1}^{(1)}$ & $16.38 \pm 0.05$ & $6.9 \pm 1.0$ & $6.9 \pm 2.0$ &
$16.35 \pm 0.09$ & $1.3 \pm 0.6$ & $0.9 \pm 0.7$ \\ 

- & $18.24 \pm 0.12$ & $0.8 \pm 0.6$ & $0.8 \pm 1.6$ & 
$18.05 \pm 0.18$ & $0.3 \pm 0.5$ & -  \\ 

$\Psi_{1}^{(2)}$ & $20.53 \pm 0.08$ & $4.7 \pm 1.0$ & $2.1 \pm 1.8 $  &
$20.37 \pm 0.01$ & $21.0 \pm 0.8$ & $20.0 \pm 0.8$ \\ 

$\Psi_{1}^{(1)}$ & $22.32 \pm 0.02$ & $8.4 \pm 0.9$ & $4.1 \pm 1.6$ &
$22.31 \pm 0.02$ & $10.5 \pm 0.8$ & $13.9 \pm 0.8$ \\ 

$\Phi_{2}^{(1,2)}$ & $27.58 \pm 0.04$ & $14.3 \pm 1.3$ & $9.1 \pm 2.1 $ &
$27.67 \pm 0.02$ & $23.5 \pm 0.7$ & $11.5 \pm 1.0$ \\ 
\hline 
\end{tabular} 
\end{table*}

\end{document}